\documentclass[10pt,twocolumn,english]{IEEEtran}
\usepackage[T1]{fontenc}
\usepackage[latin9]{inputenc}
\usepackage{amsthm}
\usepackage{amsmath}
\usepackage{amssymb}
\usepackage{graphicx}

\makeatletter
\theoremstyle{plain}
\newtheorem{thm}{\protect\theoremname}

\usepackage{amsthm}

\theoremstyle{plain}

\theoremstyle{remark}

\theoremstyle{plain}

\theoremstyle{plain}

\ifCLASSINFOpdf
\else
\fi
\usepackage{babel}

\usepackage{babel}

\providecommand{\theoremname}{Theorem}

\usepackage{babel}
\providecommand{\theoremname}{Theorem}

\usepackage{babel}
\providecommand{\theoremname}{Theorem}

\makeatother

\usepackage{babel}
\providecommand{\theoremname}{Theorem}

\begin{document}

\title{On the Transfer of Information and Energy in Multi-User Systems}

\author{Ali Mohammad Fouladgar, \IEEEmembership{Student Member,~IEEE,} and
Osvaldo Simeone,~\IEEEmembership{Member,~IEEE} 
\thanks{A. M. Fouladgar and O. Simeone are with the Center for Wireless Communications
and Signal Processing Research (CWCSPR), ECE Department, New Jersey
Institute of Technology (NJIT), Newark, NJ 07102, USA (email: af82@njit.edu,
osvaldo.simeone@njit.edu). %
}%
\thanks{The work of O. Simeone was supported in part by the U.S. National
Science Foundation under Grant No. 0914899. %
}}
\maketitle
\begin{abstract}
The problem of joint transfer of information and energy for wireless
links has been recently investigated in light of emerging applications
such as RFID and body area networks. Specifically, recent work has
shown that the additional requirements of providing sufficient energy
to the receiver significantly affects the design of the optimal communication
strategy. In contrast to most previous works, this letter focuses
on baseline \emph{multi-user} systems, namely multiple access and
multi-hop channels, and demonstrates that energy transfer constraints
call for additional coordination among distributed nodes of a wireless
network. The analysis is carried out using information-theoretic tools,
and specific examples are worked out to illustrate the main conclusions.\end{abstract}

\section{Introduction}

Electromagnetic waves carry both energy and information. Information
is modulated on the amplitude and phase of an electromagnetic wave,
while energy transfer is realized via either near-field induction
or far-field radiation. Applications of wireless energy transfer include
passive radio-frequency identification (RFID) \cite{chawla}, body
area networks \cite{zhang et al}, and power beaming using microwaves
or laser from satellites or aircraft \cite{matsumoto}.

Recent research has recognized that the two tasks of energy and information
pose conflicting constraints on the design of a wireless system \cite{varshney-1}-\cite{varshney1}.
This can be easily understood by noting that the power of a signal
depends on its average squared value, while the quantity of information
depends on the amount of ``variations'', or more specifically on
the entropy rate, of the signal itself -- maximizing one generally
does not lead to a maximum of the other.

Previous work \cite{varshney-1}-\cite{varshney1} has focused on
\emph{point-to-point }\textit{\emph{or}}\emph{ broadcast} systems,
and specifically on the problem of maximizing the information rate
subject to minimum received energy constraints. Reference \cite{varshney-1}
studied a single point-to-point channel, \cite{grover,varshney1}
investigated a set of parallel point-to-point channels and \cite{zhang}
considered a multi-antenna broadcast channel. It was shown that, as
the argument above suggests, the design of the optimal transmission
strategy depends drastically on the received requirements. 
\begin{figure}[h!]
\centering \includegraphics[bb=2bp 465bp 455bp 606bp,clip,scale=0.45]{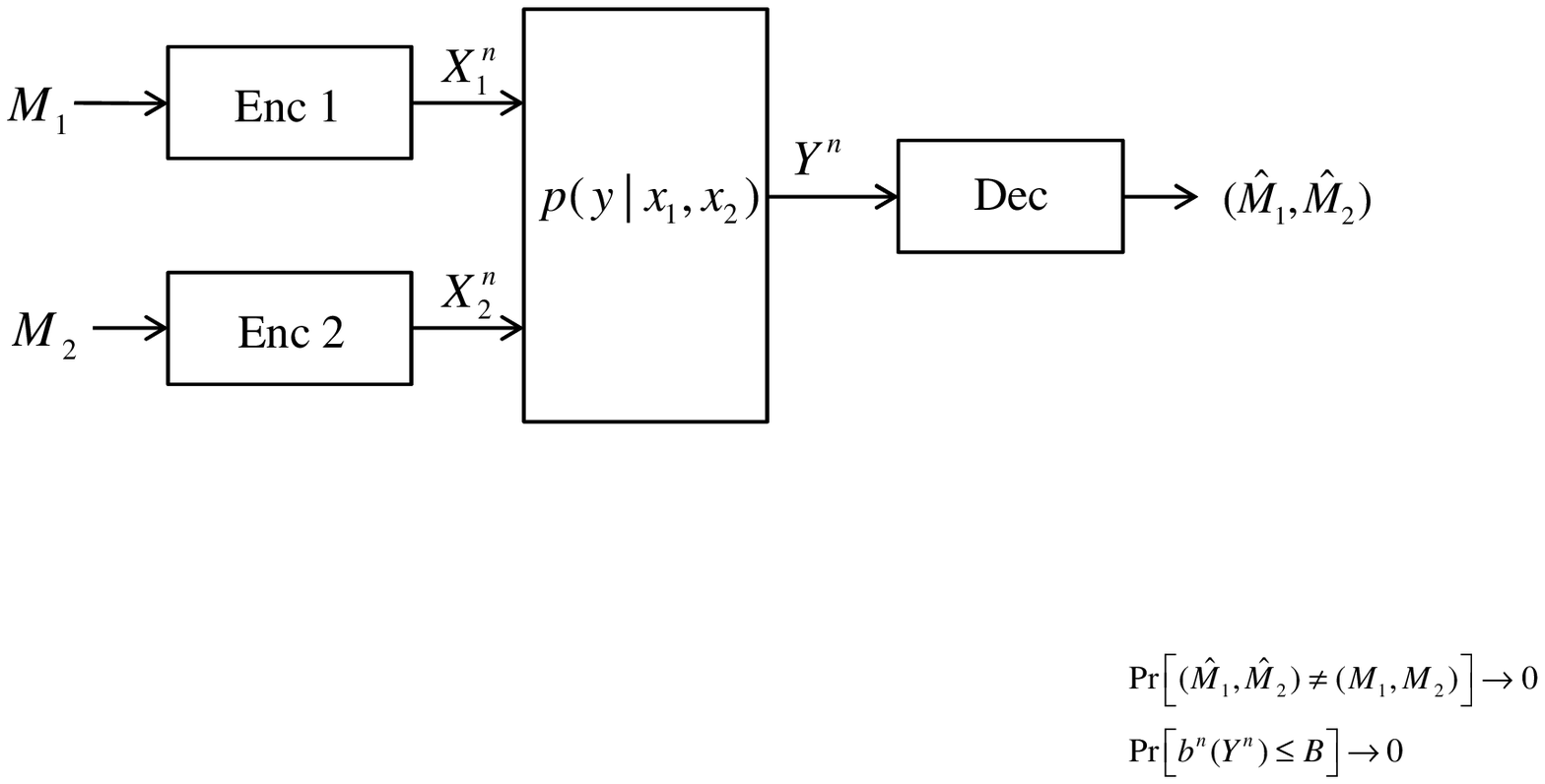}
\caption{DM-MAC with independent messages and received energy constraints.}
\label{fig:fig1} 
\end{figure}
 Incidentally, we note that the setting at hand leads to very different
insights with respect to the problem of maximizing the information
rate subject to \emph{maximum} received energy constraints considered
in \cite{gastpar}%
\footnote{As an instance of this fact, one can compare \cite[Theorem 1]{grover}
with \cite[Theorem 2]{gastpar}.%
}. We also remark that the model considered in \cite{gurakan} is different
in that it assumes that energy can be transferred between two nodes
via a separate channel devoted to energy transfer.

\begin{figure}
\centering\includegraphics[bb=2bp 477bp 509bp 519bp,clip,scale=0.45]{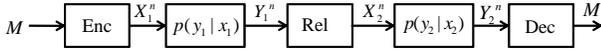}

\caption{DM-MHC with a relay that can harvest the received energy.}

\label{fig:fig2} 
\end{figure}

While previous work focused on systems with a single transmitter,
in this letter we take a first look at systems with multiple transmitters
by focusing on the two baseline scenarios of \textit{multiple access
channels} (Fig. \ref{fig:fig1}) and \textit{multi-hop channels} (Fig.
\ref{fig:fig2}). The main aim of this study is to argue that novel
forms of coordination among distributed transmitters of a wireless
networks become useful when the system design has to account for energy
transfer requirements. More specifically, our contributions are as
follows. 
\begin{itemize}
\item \textbf{Multiple access channel with received energy constraint: }In
Sec. \ref{sec:Multiple-Access-Channel}, we consider the standard
multiple access channel in Fig. \ref{fig:fig1} with the additional
constraint that the energy received by the decoder be large enough.
First, the characterization of all the achievable trade-off among
rate pairs and received energy is obtained, extending the point-to-point
result of \cite{varshney-1}. Then, an example is provided that demonstrates
the enhanced need for coordination between the two encoders in order
to satisfy the requirement on energy transfer; 
\item \textbf{Multi-hop channel with a harvesting relay: }In Sec. \ref{sec:Multi-Hop-Channel-with},
we turn our attention to the multi-hop channel in Fig. \ref{fig:fig2},
where the relay is assumed to be able to harvest the energy received
from the encoder for transmission to the decoder. A characterization
of the capacity is derived. An example is then described that illustrates
the novel issues that arise in the design of the communication strategy
in the first hop due to the harvesting capabilities for transmission
over the second hop.
\end{itemize}

\section{Multiple Access Channel With Received Energy Constraint\label{sec:Multiple-Access-Channel}}

In this section, we consider a Discrete Memoryless Multiple Access
Channel (DM-MAC) ($\mathcal{X}_{1}\times\mathcal{X}_{2},p(y|x_{1},x_{2}),\mathcal{Y})$
in which two encoders wish to communicate independent messages to
the decoder and at the same time, to provide the latter with sufficient
received energy (see Fig. \ref{fig:fig1}). The channel is described
by three finite alphabets $\mathcal{X}_{1},\mathcal{X}_{2},\mathcal{Y}$,
which are subsets of real numbers, and a collection of conditional
probability mass functions (pmfs) $p(y|x_{1},x_{2})$ on $\mathcal{Y}$.
All definitions are standard, see, e.g., \cite[Chapter 4]{El Gamal},
except for the requirement on the received energy to be discussed
below. We use the same notation as \cite{El Gamal}.

Specifically, a $(2^{nR_{1}},2^{nR_{2}},P_{1},P_{2},n)$ code for
the DM-MAC consists of: 
\begin{itemize}
\item two message sets%
\footnote{$[1:n]=\{1,...,n\}$ for any integer $n$.%
} $[1:2^{nR_{1}}]$ and $[1:2^{nR_{2}}]$; 
\item two encoders, where encoder 1 assigns a codeword%
\footnote{We denote $X^{n}$ as the sequence $X^{n}=[X_{1},...,X_{n}]$.%
} $x_{1}^{n}(m_{1})$ to each message $m_{1}\in[1:2^{nR_{1}}]$ and
encoder 2 assigns a codeword $x_{2}^{n}(m_{2})$ to each message $m_{2}\in[1:2^{nR_{2}}]$.
We have the input cost constraints 
\begin{equation}
c_{k}^{n}(x_{k}^{n}(m_{k}))=\frac{1}{n}\overset{n}{\underset{i=1}{\sum}}c_{k}(x_{ki}(m_{k}))\leq P_{k}
\end{equation}
 for given functions $c_{k}:\mathcal{X}\rightarrow\mathbb{R}_{+}$,
for all messages $m_{k}\in[1:2^{nR_{k}}]$, and $k=1,2;$ 
\item a decoder that assigns an estimate $(\hat{m}_{1},\hat{m_{2}})\in[1:2^{nR_{1}}]$$\times[1:2^{nR_{2}}]$
to each received sequence $y^{n}$. 
\end{itemize}
The message pair ($M_{1},M_{2})$ is uniformly distributed in the
set $[1:2^{nR_{1}}]\times[1:2^{nR_{2}}]$. The average probability
of error is defined as $P_{e}^{(n)}=\textrm{Pr}\{(\hat{M_{1}},\hat{M_{2}})\neq(M_{1},M_{2})\}$.
The\textit{ received energy} for a sequence $y^{n}$ is defined as
\begin{equation}
b^{n}(y^{n})=\frac{1}{n}\overset{n}{\underset{i=1}{\sum}}b(y_{i}),\label{eq:Energy transfer}
\end{equation}
 for a given function $b:\mathcal{Y}\rightarrow\mathbb{R}_{+}$. A
rate-energy triple $(R_{1},R_{2},B)$ is said to be \textit{achievable}
with energy cost constraints $(P_{1},P_{2})$ for the DM-MAC if there
exists a sequence of $(2^{nR_{1}},2^{nR_{1}},P_{1},P_{2},n)$ codes
such that 
\begin{equation}
\limsup_{n\rightarrow\infty}P_{e}^{(n)}=0\label{eq:2a}
\end{equation}
 and 
\begin{equation}
\limsup_{n\rightarrow\infty}\textrm{Pr}\left[\frac{1}{n}\overset{n}{\underset{i=1}{\sum}}b(Y_{i})<B-\epsilon\right]=0.\label{eq:Bbound}
\end{equation}
 for any $\epsilon>0$. Condition (\ref{eq:Bbound}) states that the
average received energy should be at least $B$ with high probability
as $n$ grows large%
\footnote{This entails also the weaker constraint $1/n\sum_{i=1}^{n}\mathrm{E}[b(Y_{i})]\geq B-\epsilon$,
for $n$ large enough (see Appendix A).%
}. The \textit{capacity-energy region} $\mathcal{C}_{e}(P_{1},P_{2})$
of the DM-MAC is the closure of the set of achievable rate-energy
triple $(R_{1},R_{2},B)$.

\subsection{Capacity-Energy Region }

We are now ready to present a characterization of the DM-MAC capacity-energy
region. 
\begin{thm}
The capacity-energy region $\mathcal{C}_{e}(P_{1},P_{2})$ of the
DM-MAC with received energy constraint is the union of the set of
all rate-energy triples $(R_{1},R_{2},B)$ such that the inequalities
\begin{subequations}\label{eq:prop1} 
\begin{eqnarray}
R_{1} & \leq & I(X_{1};Y|X_{2},Q),\\
R_{2} & \leq & I(X_{2};Y|X_{1},Q),\\
R_{1}+R_{2} & \leq & I(X_{1},X_{2};Y|Q),\\
\textrm{and }B & \leq & \textrm{E}[b(Y)]
\end{eqnarray}
 \end{subequations}hold for some pmfs $p(q),\textrm{ }p(x_{1}|q)$
and $p(x_{2}|q)$ satisfying the constraints $\textrm{E}[c(X_{k})]\leq P_{k},\textrm{ for }k=1,2$.
The alphabet of $Q$ can be bounded as $|\mathcal{Q}|\leq4$. 
\end{thm}
It is recalled, for future reference, that the variable $Q$ in (\ref{eq:prop1})
enables time-sharing between different pairs of codebooks used by
the two encoders. Specifically, when $Q=q$ for some $q\in\mathcal{Q}$,
the codebooks to be used by the encoders are characterized by the
conditional pmfs $p(x_{1}|q)$ and $p(x_{2}|q)$ as per standard random
coding arguments. We remark that time-sharing requires coordination
between the encoders that have to agree on a sequence $Q^{n}$ and
switch to the preassigned codebooks when appropriate. 
\begin{IEEEproof}
Achievability and converse follow with minor variations from standard
results on the DM-MAC (see \cite[Chapter 4]{El Gamal}). Specifically,
achievability follows from the same random coding arguments in \cite[Chapter 4]{El Gamal}.
The only difference is that, in order to guarantee the constraint
(\ref{eq:Bbound}), an additional error event is added to the conventional
analysis in \cite[Chapter 4]{El Gamal} for the case in which the
event within square brackets in (\ref{eq:Bbound}) does not hold.
This event is immediately shown to have vanishing probability for
large $n$ by the weak law of large numbers. Some details on the converse
can be found in Appendix A.

\begin{figure}
\centering\includegraphics[bb=70bp 333bp 455bp 620bp,clip,scale=0.45]{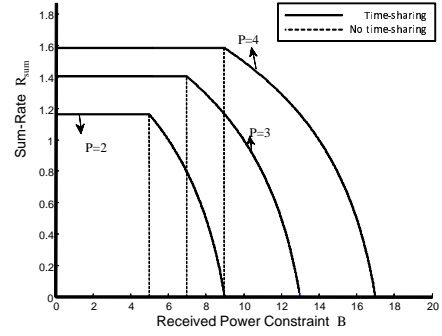}

\caption{Sum-rate $R_{sum}$versus received energy constraint $B$ for the
Gaussian MAC studied in Sec. \ref{sub:Example:-Gaussian-MAC}.}

\label{fig:fig3} 
\end{figure}

\end{IEEEproof}

\subsection{Example\label{sub:Example:-Gaussian-MAC}}

In order to illustrate the novel aspects in the system design that
are entailed by the minimum receive energy constraint (\ref{eq:Bbound}),
we now consider the Gaussian MAC $Y=X_{1}+X_{2}+Z,$ where $Z\sim\mathcal{N}(0,1)$
is the additive noise. Let the input cost function be $c_{k}(x_{k})=x_{k}^{2}$
for $k=1,2$ and the received energy function be $b(y)=y^{2}$. Note
that the Gaussian MAC is not DM, but the result in Theorem 1 applies
as per standard arguments \cite[Sec. 3.4]{El Gamal}. From Theorem
1, all achievable sum-rates for the Gaussian MAC at hand can be written
as 
\begin{equation}
R_{sum}=I(X_{1},X_{2};Y|Q),\label{eq:optRsum}
\end{equation}
 for some pmfs $p(q),p(x_{1}|q)$ and $p(x_{2}|q)$ under the constraints
$\textrm{E}[Y^{2}]\geq B$ and $\textrm{E}[X_{k}^{2}]\leq P_{k},\textrm{ for }k=1,2$.

As it is well known (see, e.g., \cite{El Gamal}), without any constraint
on the received energy, the maximum sum-rate is given as 
\begin{eqnarray}
R_{sum} & \negmedspace\negmedspace\negmedspace=\negmedspace\negmedspace\negmedspace & \frac{1}{2}\log_{2}(1+2P),\label{eq:maxgaussianrate}
\end{eqnarray}
 which is achieved by setting $Q$ to a constant and $X_{k}\sim\mathcal{N}(0,P)$
for $k=1,2$. In other words, \emph{maximum information transfer does
not require time sharing}. Moreover, with this choice, the received
power is $E[Y^{2}]=2P+1$. Therefore, if $B\leq2P+1$, then, even
under the received energy constraint (\ref{eq:Bbound}), the maximum
sum-rate is given by (\ref{eq:maxgaussianrate}) and time sharing
is not needed.

Now we assume that $B>2P+1$, and consider a time-sharing strategy
whereby $Q\sim Ber(\lambda)$ and $X_{1},X_{2}\sim\mathcal{N}(0,P^{\prime})$
for $Q=1$ and $X_{1}=X_{2}=\sqrt{P^{\prime\prime}}$ for $Q=0$ for
some $0\leq\lambda\leq1\textrm{, }P^{\prime}\geq0$ and $P^{\prime\prime}\geq0$.
In other words, for $Q=1$ information-maximizing codebooks are used,
while for $Q=0$ the two encoders transmit the constant signals $X_{1}=X_{2}=\sqrt{P^{\prime\prime}}$.
The latter choice maximizes the energy transfer due to coherent combining
at the receiver, but carries no information. Optimizing over the parameters
$(P^{\prime},P^{\prime\prime},\lambda)$, from (\ref{eq:optRsum}),
the following sum-rate is achievable 
\begin{equation}
\begin{array}{ccc}
R_{sum} & =\negmedspace\negmedspace\underset{0\leq\lambda\leq1,\textrm{ }P^{\prime\prime},P^{\prime\prime}\geq0}{\textrm{maximize}}\negmedspace\negmedspace\negmedspace\negmedspace\negmedspace\negmedspace & \negmedspace\negmedspace\negmedspace\negmedspace\negmedspace\negmedspace\frac{\lambda}{2}\log_{2}(1+2P^{\prime})\\
 & \textrm{\textrm{subject to}}\negmedspace\negmedspace\negmedspace\negmedspace\negmedspace\negmedspace & \negmedspace\negmedspace\negmedspace\negmedspace\negmedspace\negmedspace\lambda P^{\prime}+(1-\lambda)P^{\prime\prime}\leq P,\\
 & \negmedspace\negmedspace\negmedspace\negmedspace\negmedspace\negmedspace & \negmedspace\negmedspace\negmedspace\negmedspace\negmedspace\negmedspace B\leq2\lambda P^{\prime}+4(1-\lambda)P^{\prime\prime}+1.
\end{array}
\end{equation}
 This rate is shown in Figure \ref{fig:fig3} versus the constraint
on the received energy $B$ for different values of $P$ along with
the sum-rate (\ref{eq:maxgaussianrate}) obtained with no time-sharing.
It is seen that, as the received energy constraint $B$ becomes large
enough (i.e., $B>2P+1$), \emph{time-sharing is necessary to achieve
the optimal performance}. This demonstrates that additional coordination
is generally needed between the encoders in order to obtain the desired
trade-off between information and energy transfer.

\section{Multi-Hop Channel with a Harvesting Relay\label{sec:Multi-Hop-Channel-with}}

In this section, we consider the three-node Discrete Memoryless Multi-Hop
Channel (DM-MHC) in Figure \ref{fig:fig2}, in which the encoder wishes
to communicate a message $M$ to the decoder with the help of a relay.
The relay can harvest the energy received from the encoder as formalized
below. We refer to the relay as having energy-harvesting capabilities.
The DM-MHC is characterized by two separate DM point-to-point channels
$(\mathcal{X}_{1},p(y_{1}|x_{1}),\mathcal{Y}_{1})$ and $(\mathcal{X}_{2},p(y_{2}|x_{2}),\mathcal{Y}_{2})$.
All definitions are standard, see, e.g., \cite[Chapter 16]{El Gamal},
except for the fact that the relay can harvest energy from the received
signal.

Specifically, a $(2^{nR},P_{1},P_{2},n)$ code for the DM-MHC consists
of 
\begin{itemize}
\item a message set $\left[1:2^{nR}\right]$; 
\item an encoder that assigns a codeword $x_{1}^{n}(m)$ to each message
$m\in\left[1:2^{nR}\right]$. We have the \textit{input cost constraint}
\begin{equation}
c_{1}^{n}(x_{1}^{n}(m))=\frac{1}{n}\overset{n}{\underset{i=1}{\sum}}c_{1}(x_{1i}(m))\leq P_{1}\label{eq:14powercost}
\end{equation}
 for some function $c_{1}:\mathcal{X}_{1}\rightarrow\mathbb{R}_{+}$,
and for all messages $m\in\left[1:2^{nR}\right]$; 
\item a relay encoder that assigns a symbol $x_{2i}(y_{1}^{i-1})$ to each
past received sequence $y_{1}^{i-1}\in\mathcal{Y}_{1}^{i-1}$ for
each time $i\in\left[1:n\right]$. The relay input is constrained
so that condition 
\begin{equation}
\begin{array}{ccc}
\frac{1}{n}\overset{n}{\underset{i=1}{\sum}}c_{2}(x_{2i}) & \leq & \frac{1}{n}\overset{n}{\underset{i=1}{\sum}}b(y_{1i})\end{array}+P_{2}\label{eq:15Energy-cost}
\end{equation}
 is satisfied. This implies that the input cost $\frac{1}{n}\sum_{i=1}^{n}c_{2}(x_{2i})$
should be less than the average harvested energy $\frac{1}{n}\sum_{i=1}^{n}b(y_{1i})$
and the available power $P_{2}$; 
\item a decoder that assigns an estimate $\hat{m}\in\left[1:2^{nR}\right]$
to each received sequence $y_{2}^{n}$. 
\end{itemize}
We assume that the message $M$ is uniformly distributed in the set
$\left[1:2^{nR}\right]$. The average probability of error is defined
as $P_{e}^{(n)}=\textrm{Pr}\{\hat{M}\neq M\}$. A rate $R$ is said
to be \textit{achievable} with energy cost constraints $P_{1},P_{2}$
if there exists a sequence of $(2^{nR},P_{1},P_{2},n)$ codes such
that (\ref{eq:2a}) is satisfied. The \textit{capacity-energy function}
$\mathcal{C}_{e}(P_{1},P_{2})$ of the DM-MHC is the supremum of the
set of all achievable rates $R.$

\subsection{Capacity-Energy Function}

We now present a characterization of the DM-MHC capacity-energy function. 
\begin{thm}
The capacity-energy function $\mathcal{C}_{e}(P_{1},P_{2})$ of the
DM-MHC with a harvesting relay is 
\begin{eqnarray}
\mathcal{C}_{e}(P_{1},P_{2})\negmedspace\negmedspace & \negmedspace\negmedspace=\negmedspace\negmedspace\negmedspace\negmedspace & \underset{p(x_{1}):\textrm{ E}[c_{1}(X_{1})]\leq P_{1}}{\max}\min\negmedspace\negmedspace\left\{ \underset{}{I(X_{1};Y_{1}),\negmedspace\negmedspace}\right.\nonumber \\
 & \negmedspace\negmedspace\negmedspace\negmedspace & \left.\negmedspace\negmedspace\underset{p(x_{2}):\textrm{ E}[c_{2}(X_{2})]\leq\textrm{E}[b(Y_{1})]+P_{2}}{\max}\negmedspace\negmedspace I(X_{2};Y_{2})\right\} \negmedspace.
\end{eqnarray}
 \end{thm}
\begin{IEEEproof}
The achievability follows via decode-and-forward using the same arguments
as in \cite[Ch. 16]{El Gamal}. The only difference is that, in order
to guarantee that condition (\ref{eq:15Energy-cost}) is satisfied,
an error event is introduced for the case where (\ref{eq:15Energy-cost})
does not hold, that is shown to have vanishing probability as $n\rightarrow\infty$
by the weak law of large numbers. Some details on the converse can
be found in Appendix B. 
\end{IEEEproof}

\subsection{Example}

Consider a noiseless channel $Y_{1}=X_{1}$ with $\mathcal{X}_{1}=\mathcal{Y}_{1}=\{-2,-1,1,2\}$
followed by a Gaussian channel $Y_{2}=X_{2}+Z,$ with $Z\sim\mathcal{N}(0,N_{0})$.
Note that the second channel is not DM, but the analysis applies using
conventional arguments \cite[Sec. 3.4]{El Gamal}. Let the input cost
functions be $c_{k}(x_{k})=x_{k}^{2}$ for $k=1,2$ and the received
energy function be $b(y_{1})=y_{1}^{2}$.

We first observe that, if $b(y_{1})=0$ for all $y_{1}\in\{-2,-1,1,2\}$
and thus no energy can be harvested at the relay, then the capacity
is given by $\textrm{min }\{2,1/2\log_{2}(1+P_{2})\}$, which is achieved
by setting $X_{1}$ to be uniformly distributed in the set $\{-2,-1,1,2\}$
and $X_{2}\sim\mathcal{N}(0,P_{2})$. Note that,\emph{ in this conventional
case, the codebook selected by the encoder depends on the quality
of the second link only through the rate}, which needs to be set to
$C$, but is otherwise independent since $X_{1}$ is uniformly distributed
irrespective of the quality of the second hop.

\begin{figure}
\centering\includegraphics[bb=58bp 325bp 434bp 634bp,clip,scale=0.45]{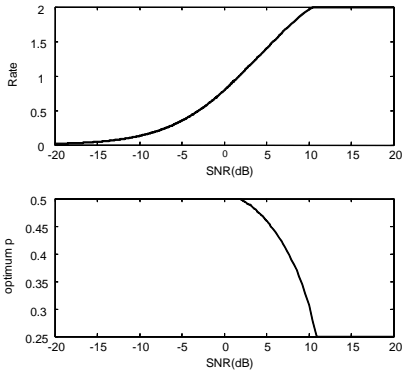}

\caption{Capacity-energy $\mathcal{C}_{e}(P_{1},P_{2})$ versus the SNR (upper
figure) and optimum probability $p$ (lower figure) for the multi-hop
channel studied in Sec. III-B ($P_{1}=4,P_{2}=0$).}
\label{fig:fig4}
\end{figure}

We now consider the effect of a harvesting relay. From Theorem 2,
and using the symmetry of the problem, the capacity-energy function
can be written as 
\begin{eqnarray}
\negmedspace\negmedspace\negmedspace\negmedspace\negmedspace\negmedspace\negmedspace\negmedspace\mathcal{C}_{e}(P_{1},P_{2}) & \negmedspace\negmedspace\negmedspace\negmedspace=\negmedspace\negmedspace\negmedspace\negmedspace\negmedspace\negmedspace & \underset{6p+1\leq P_{1}\negmedspace\negmedspace\negmedspace\negmedspace}{\underset{0\leq p\leq1/2:\negmedspace\negmedspace\negmedspace\negmedspace}{\textrm{max}}}\textrm{min\negmedspace}\left\{ \negmedspace-2p\textrm{log}_{2}p-\negmedspace\negmedspace(1-2p)\textrm{log}_{2}\negmedspace\left(\frac{1}{2}-p\right)\negmedspace,\negmedspace\negmedspace\negmedspace\negmedspace\negmedspace\negmedspace\negmedspace\textrm{ }\right.\nonumber \\
 & \negmedspace\negmedspace\negmedspace\negmedspace & \left.\negmedspace\negmedspace\negmedspace\negmedspace\frac{1}{2}\textrm{log}_{2}\left(1+\frac{P_{2}+6p+1}{N_{0}}\right)\negmedspace\right\} \negmedspace.\negmedspace\negmedspace\negmedspace\negmedspace\negmedspace\negmedspace\negmedspace\negmedspace
\end{eqnarray}
 This is obtained by setting $p=\textrm{Pr}[X_{1}=2]=\textrm{Pr}[X_{1}=-2]$
and $1/2-p=\textrm{Pr}[X_{1}=1]=\textrm{Pr}[X_{1}=-1]$ and $X_{2}\sim\mathcal{N}(0,P_{2}+\textrm{E}[Y_{1}^{2}])$
given that $\textrm{E}[Y_{1}^{2}]=\textrm{E}[X_{1}^{2}]=2\left(4p+(\frac{1}{2}-p)\right)=6p+1$.
The capacity and the optimum value of $p$ are shown in Figure \ref{fig:fig4}
versus the signal-to-noise ratio SNR=$10\textrm{log}_{2}\left(1/N_{0}\right)$
for $P_{1}=4,$ $P_{2}=0$.

As it can be seen, for small SNRs in the second hop, it is advantageous
to maximize the energy transfer to the relay, which is obtained for
$p=0.5$. Instead, for sufficiently large SNR in the second hop, it
is optimal to maximize the information transfer to the relay, which
is obtained for $p=0.25$. \textit{T}\emph{his demonstrates that,
in a multi-hop channel with a harvesting relay, the encoder needs
to fully adjust its transmission strategy depending on the quality
of the second link}. This calls for a larger degree of coordination
than in the conventional scenario.

\section{Conclusions}

The two baseline multi-user scenarios studied in this paper complement
the initial work \cite{varshney-1}-\cite{varshney1} by showing that
the requirements of energy and information flow have significant consequences
on the design of wireless networks with multiple terminals. Recent
work has reached related conclusions for a two-way communication model
\cite{popovski}. An interesting avenue for future work is the design
of practical coding strategies that achieve a desired trade-off between
energy and information transfer.

\appendices{}

\section*{Appendix A: Sketch of Converse Proof for Theorem 1}

The proof of the converse follows the same procedure as in \cite[page 89]{El Gamal},
except for the proof of the inequality $B\leq E[b(Y)]$. To prove
this bound, note that any $(2^{nR_{1}},2^{nR_{1}},P_{1},P_{2},n)$
code needs to satisfy (\ref{eq:Bbound}). Moreover, we have the bound
\begin{eqnarray}
\textrm{E}\left[b^{n}(Y^{n})\right] & \negmedspace\negmedspace\geq\negmedspace\negmedspace\negmedspace & (B-\epsilon)\textrm{Pr}\left[b^{n}(Y^{n})\geq B-\epsilon\right],
\end{eqnarray}
 and thus, by (\ref{eq:Bbound}), we also have that $\textrm{E}\left[b^{n}(Y^{n})\right]\geq B-\epsilon$
for $n$ large enough. Given that $\textrm{E}\left[b^{n}(Y^{n})\right]=\frac{1}{n}\sum_{i=1}^{n}\textrm{E}\left[b(Y_{i})\right]$,
defining $Y$ as in \cite[page 81]{El Gamal} concludes the proof.

\section*{Appendix B: Sketch of Converse Proof for Theorem 1}

From the cut-set bound in \cite[Theorem 16.1]{El Gamal}, we have
the inequality 
\begin{eqnarray}
R & \negmedspace\negmedspace\leq\negmedspace\negmedspace & \underset{p(x_{1},x_{2})}{\max}\min\left\{ I(X_{1};Y_{1}),I(X_{2};Y_{2})\right\} .\label{eq:26}
\end{eqnarray}
 We recall that this is proved by defining $Q\sim\textrm{Unif}\left[1:n\right]$
independent of $(X_{1}^{n},X_{2}^{n})$ and by setting $X_{1}=X_{1Q}$,
$X_{2}=X_{2Q}$. With these definitions, from (\ref{eq:14powercost})
and (\ref{eq:15Energy-cost}), we get $\textrm{ E}[c_{1}(X_{1})]\leq P_{1}$
and $\textrm{ E}[c_{2}(X_{2})]\leq\textrm{E}[b(Y_{1})]+P_{2}$, respectively.
We then observe that the two terms inside the min function are separately
functions of $p(x_{1})$and $p(x_{2})$, so that we can maximize over
the marginals $p(x_{1})$ and $p(x_{1})$. The order of the optimizations
in (\ref{eq:26}) follows the same arguments.

\end{document}